\documentclass[aps]{revtex4}
\usepackage{eurosym}
\usepackage{amsfonts}
\usepackage{amsmath}
\usepackage{amssymb,epsf}
\usepackage{color}
\usepackage{graphicx}
\usepackage{natbib}
\usepackage{float}
\usepackage{caption}
\usepackage{subfigure}
\usepackage{epstopdf}

\begin{document}

 \title{Stability and phase transition of rotating Kaluza-Klein black holes}
 \author{
 Seyed Hossein Hendi$^{1,2,3}$ \footnote{hendi@shirazu.ac.ir (corresponding author)},
 Somayeh Hajkhalili$^{1,2}$\footnote{Hajkhalili@gmail.com},
 Mubasher Jamil$^{4,5,3}$\footnote{mjamil@zjut.edu.cn (corresponding author)} and
 Mehrab Momennia$^{1,2,6}$\footnote{momennia1988@gmail.com}}

 \affiliation{
 $^1$Department of Physics, School of Science, Shiraz University, Shiraz 71454, Iran \\
 $^2$Biruni Observatory, School of Science, Shiraz University, Shiraz 71454, Iran \\
 $^3$Canadian Quantum Research Center 204-3002 32 Ave Vernon, BC V1T 2L7 Canada \\
 $^4$Institute for Theoretical Physics and Cosmology, Zhejiang University of Technology, Hangzhou, 310023 China \\
 $^5$School of Natural Sciences, National University of Sciences and Technology, H-12, Islamabad 44000, Pakistan\\
 $^6$Instituto de F\'{i}sica, Benem\'{e}rita Universidad Aut\'{o}noma de Puebla. Apdo. Postal J-48, C.P. 72570, Puebla, M\'{e}xico}

\begin{abstract}
 In this paper, we investigate thermodynamics and phase transitions of a 4-dimensional rotating Kaluza-Klein black hole solution in the presence
 of Maxwell electrodynamics. Calculating the conserved and thermodynamical quantities shows that the first law of thermodynamics is satisfied.
 To find the stable black hole's criteria, we check the stability in the canonical ensemble by analyzing the behavior of the heat capacity.
 We also consider a massive scalar perturbation minimally coupled to the background geometry of the $4$-dimensional static Kaluza-Klein black hole
 and investigate the quasinormal modes by employing the WKB approximation. The anomalous decay rate of the quasinormal modes spectrum is investigated
 by using the sixth-order WKB formula and quasi-resonance modes of the black hole are studied with averaging of Pad\'{e} approximations as well.
\end{abstract}

\maketitle


\section{Introduction}

The Kaluza Klein (KK) theory is one of the oldest theories of the last
century which proposed the extension of general relativity in higher
dimensions and incorporated extra fields such as electromagnetism. In the
past few decades, the KK theory has been extended into a more general class
of string theories, however, the KK theory is still relevant as a low energy
effective version of string theory \cite{KK,KK1,KK2}. The simplest KK theory
is obtained using the general relativity in five dimensions and then
subsequently dimensionally reducing to four dimensions. This extended
framework contains both gravity and electromagnetism. Although the original
five dimensional theory is not a realistic theory of nature, it has been
interpreted in a quantum mechanical framework as well as string theory. KK
theory has also been attracted to non-commutative differential geometry
which may be viewed as KK theory in which the extra fifth dimension is taken
to be a discrete set of points rather than a continuum \cite{KK7}. However,
stationary spherically symmetric BHs were derived in \cite{KK5}. Rotating
black hole (BH) solutions in four and five dimensions were found by Larsen
\cite{KK4}. These BHs typically have four free parameters: mass, spin, the
electric and magnetic charges. Also, six and higher dimension versions of KK
BHs have been constructed \cite{KK6}. In this paper, we study a four
dimensional rotating KK BH.

From thermodynamic considerations, we know that under certain
conditions, a thermodynamic system can experience a phase
transition. Davies found that a discontinuity of the heat capacity
represents the second order phase transition in BHs about forty
years ago \cite{phase1} while the analysis of phase structure was
done by Hut \cite{phase2}. This approach is in the framework of
the canonical ensemble. Note that in the extremal limit heat
capacity vanishes which occurs because of zero temperature. It is
called type one phase transition since the sign of heat capacity
is changed, hence the BH with negative heat capacity is unstable,
and therefore, the system is undergoing a phase transition.



When the background spacetime of a BH undergoes dynamical perturbations, the
resultant behavior involves some sort of oscillations in spacetime geometry
called quasinormal modes (QNMs). The QNMs are independent of initial
perturbed configuration and they are the intrinsic fingerprint of the BH
response to the external perturbations. The QNMs usually have an imaginary
part giving the damping time of perturbations while a real part representing
the actual oscillations. Investigating vibrations in the background geometry
of BHs is one of the most important and exciting topics in the context of
compact stars physics and these oscillations describe the evolution of
fields on the background spacetime \cite{Kokkotas,BertiRev}. The QNM
spectrum reflects the properties of the spacetime and we can probe the
properties of the background by studying these vibrations. Therefore, the
perturbed BH encodes its intrinsic properties, such as mass, charge, and
angular momentum in the QNM spectrum. The QNMs of supermassive BHs
undergoing gravitational perturbations can be observed by future space-based
gravitational wave detectors \cite{GWsupMass}, and investigation of the BH
oscillations attracted much attention recently after the detection of the
gravitational waves produced by compact binary mergers \cite{Abbott1}.

Scalar fields have been widely studied in the area of cosmology as inflatons
\cite{Cheung}, and also, considered as candidates for dark energy \cite%
{Gubitosi} and dark matter \cite{Hu}. They can play a role in constructing a
consistent theory of quantum gravity \cite{Metsaev,Arvanitaki} and modifying
the background geometry of BHs in the strong-field regime \cite%
{Herdeiro,Silva}. Scalar fields can produce clouds through instabilities
around BHs \cite{Brito,Clough}. In different models with non-minimal
interaction of scalar fields with the spacetime metric, we expect
gravitational waves to be supplemented with a scalar mode. In these models,
the gravitational waves of the spacetime geometry will be a linear
combination of gravitational waves in the underlying gravitational theory
and the scalar field solutions \cite{Tattersall}. The final QNMs are
included components oscillating with a combination of the background metric
and scalar field that could potentially be observed. Therefore, the
fingerprint of scalar fields on gravitational waves could be detected by
employing future gravitational wave detectors. However, the interactions of
scalar and tensor waves generally depend on the scalar propagation speed
such that the interactions are negligible whether the scalar waves are
luminal or quasi-luminal \cite{Dalang}.

On the other hand, a minimally coupled scalar field describes the QNMs in
the area of scalar-tensor theories, and observing quasi-resonance modes and
anomalous decay rate of QN modes motivate one to investigate these models as
well. Besides, more recently it is shown that if the primary supermassive
BHs in the extreme mass ratio inspirals do not carry a significant scalar
charge, the non-minimal coupling factor vanishes which LISA still will be
able to detect and further measure scalar charge \cite{Maselli}.

The test scalar fields minimally coupled to the background metric was
investigated for Schwarzschild BH \cite{SchwQRMs,higherSchwQRM,ADR},
Reissner-Nordstr\"{o}m BH \cite{RN-QRM,Hod-rnQRM,ADRrn}, magnetized
Schwarzschild BH \cite{magnetizedSchw-QRM}, Kerr geometry \cite{KerrQRM},
BHs in Einstein-Weyl gravity \cite{EinWeylGravityQRM}, conformal Weyl BH
solutions \cite{MehrabWeylNE,MehrabWeylGPs}, and three dimensional BHs \cite%
{Grigoris}. In this paper, we focus on perturbations of minimal coupled
massive scalar fields in the background of $4$-dimensional static KK BHs to
investigate the effects of the free parameters $p$ and $q$ on the scalar QNM
spectrum. Moreover, we shall explore the quasi-resonance modes and anomalous
decay rate of QN modes for our BH case study.

The layout of the paper is as follows. The next section is devoted to
introducing the field equations and corresponding rotating KK black holes in
four dimensions. Thermodynamic quantities such as entropy, temperature,
electric and magnetic potential as well as the examination of the first law
of BH thermodynamics are studied in Sec. \ref{Thermodynamics}. The thermal
stability of the BH in the canonical ensemble is done in Sec. \ref{Stability}%
. Then, in Sec. \ref{QNMs}, dynamical perturbations are considered and QNMs
are extracted. We finish the paper with a summary and closing remarks.


\section{Field Equations, Solutions and Thermodynamics} \label{Field}

The solution of $5-$dimensional rotating KK BHs with electric charge (Q) and
magnetic charge (P) in the presence of Maxwell electrodynamics is obtained
within the framework of four dimensional Einstein-Maxwell-Dilaton gravity
\cite{KK1}. The complete action is described as
\begin{equation}
\mathcal{S}=-\int \left( \frac{\mathcal{R}}{\kappa ^{2}}+\frac{2}{3\kappa
^{2}\varphi ^{2}}\partial _{\mu }\varphi \partial ^{\mu }\varphi +\frac{1}{4}%
\varphi ^{2}F^{2}\right) \varphi \sqrt{-g}d^{4}x,  \label{action}
\end{equation}%
where $\mathcal{R}$ displayed the Ricci scalar and $\kappa ^{2}=16 \pi G$.
Also, the dilaton field is represented by $\varphi $ and $F^{2}=F_{\mu \nu
}F^{\mu \nu }$, where $F_{\mu \nu }$ is Maxwell field tensor. Field
equations could be found by variation of the action (\ref{action}) with
respect to the metric, Maxwell and dilaton fields respectively, resulting in
\begin{eqnarray}
&&G_{\mu \nu }=\frac{\kappa ^{2}\varphi ^{2}}{2}T_{\mu \nu }^{EM}-\frac{1}{%
\varphi }\left( \partial _{\mu }\partial _{\nu }-g_{\mu \nu }\square \right)
\varphi , \\
&&T_{\mu \nu }^{EM}=F_{\mu \rho }F_{\nu }^{\rho }-\frac{1}{4}g_{\mu \nu
}F^{2}, \\
&&\partial ^{\mu }F_{\mu \nu }=-3\frac{\partial ^{\mu }\varphi }{\varphi }%
F_{\mu \nu }, \\
&&\square \varphi =\frac{\kappa ^{2}\varphi ^{3}}{3}F_{\mu \nu }F^{\mu \nu }.
\end{eqnarray}%
By using a solution generating technique, Larsen obtained the following five
dimensional solution \cite{KK4}
\begin{equation}
ds_{5}^{2}=\frac{H_{2}}{H_{1}}(dy+\mathbf{A})^{2}-\frac{H_{3}}{H_{2}}(dt+%
\mathbf{B})^{2}+H_{1}\left( \frac{dr^{2}}{\Delta }+d\theta ^{2}+\frac{\Delta
}{H_{3}}\sin ^{2}\theta d\phi ^{2}\right).
\end{equation}%
The extra coordinate $y$ is assumed to be periodic with period $2\pi R_{KK}$
where $R_{KK}$ is its radius. Also, $\partial /\partial y$ is considered to
be Killing so that the 5-dimensional metric components can be functions of $%
\{t,r,\theta ,\phi \}$ only \cite{KK3}. Here we use the following
definitions
\begin{eqnarray}
&&H_{1}=r^{2}+a^{2}\cos ^{2}\theta +r(p-2m)+\frac{p}{p+q}\frac{(p-2m)(q-2m)}{%
2}-\frac{p}{2m(p+q)}\sqrt{(q^{2}-4m^{2})(p^{2}-4m^{2})}a\cos \theta , \\
&&H_{2}=r^{2}+a^{2}\cos ^{2}\theta +r(q-2m)+\frac{q}{p+q}\frac{(p-2m)(q-2m)}{%
2}-\frac{q}{2m(p+q)}\sqrt{(q^{2}-4m^{2})(p^{2}-4m^{2})}a\cos \theta , \\
&&H_{3}=r^{2}+a^{2}\cos ^{2}\theta -2mr,\qquad \qquad \Delta
=r^{2}+a^{2}-2mr, \\
&&\mathbf{A}=-\left[ 2Q\left( r+\frac{p-2m}{2}\right) +\sqrt{\frac{%
q^{3}(p^{2}-4m^{2})}{4m^{2}(p+q)}}a\cos \theta \right] \frac{dt}{H_{2}}-%
\Bigg[2P(H_{2}+a^{2}\sin ^{2}\theta )\cos \theta +\sqrt{\frac{p(q^{2}-4m^{2})%
}{4m^{2}(p+q)^{3}}}  \notag \\
&&\qquad \times \left[ (p+q)(pr-m(p-2m))+q(p^{2}-4m^{2})\right] a\sin
^{2}\theta \Bigg]\frac{d\phi }{H_{2}}, \\
&&\mathbf{B}=\sqrt{pq}\frac{(pq+4m^{2})r-m(p-2m)(q-2m)}{2m(p+q)H_{3}}a\sin
^{2}\theta d\phi.
\end{eqnarray}%
The four free parameters $m$, $p$, $q$ and $a$ are related to the physical
parameters through
\begin{eqnarray}
&&\text{Mass:}\qquad \qquad \qquad \qquad \,\,\,M=\frac{p+q}{4},
\label{mass} \\
&&\text{magnetic charge:}\qquad \,\,\,\,\,\,\,\,\,P^{2}=\frac{p(p^{2}-4m^{2})%
}{4(p+q)},  \label{magnetic charge} \\
&&\text{electric charge:}\qquad \qquad \,\,Q^{2}=\frac{q(q^{2}-4m^{2})}{%
4(p+q)},  \label{electric charge} \\
&&\text{angular momentum:}\qquad J=\frac{\sqrt{pq}(pq+4m^{2})}{4m(p+q)}a.
\label{ang}
\end{eqnarray}

It is notable that two charges $q$ and $p$ are not independent parameters
also, they can change angular momentum and the mass of BH. Also, the
definition of charges forces us to select $p$ and $q$ larger than $2m$.
Furthermore, zero electric or magnetic charge leads the angular momentum $J$
to vanish. Besides, it is clear that two horizons can be obtained by solving
$\Delta=0$, so $r_\pm=m\pm\sqrt{m^2-a^2}$ which are impressed by mass $(M)$
and both charges $(P,Q)$ based on their definitions in Eqs. (\ref{mass})--(%
\ref{ang}).

Setting $a=m$ yields Kerr like extremal limit which named "fast
rotation" by $J > QP$. In addition if $a\to 0$ and $m\to 0$ but
the ratio $a/m<1$ we get the next extremal limit for our solution
which gets $J<PQ$ or "slow rotation" \cite{KK2}.

The corresponding four dimensional BH metric after dimensional reduction is
the following \cite{KK4}
\begin{equation}
ds_4^2=-\frac{H_3}{\sqrt{H_1H_2}}(dt+\mathbf{B})^2+ \sqrt{H_1H_2}\left(\frac{%
dr^2}{\Delta} +d\theta^2+\frac{\Delta}{H_3}\sin^2\theta d\phi^2\right)
\label{Metric4d}
\end{equation}

It is interesting to introduce the dimensionless form of the solution. In
this regard we use following dimensionless parameters \cite{KK9}
\begin{eqnarray}
&&p\equiv b\,m\qquad \qquad \,q\equiv c\,m  \notag \\
&&\epsilon ^{2}=\frac{Q^{2}}{M^{2}}\qquad \quad \,\,\mu ^{2}=\frac{P^{2}}{%
M^{2}}  \notag \\
&&\alpha =\frac{a}{M}\qquad \qquad x=\frac{r}{M}
\end{eqnarray}%
so the free independent parameters are $x,M,\alpha ,b,c$ and the metric
functions transform as
\begin{eqnarray}
\frac{H_{1}}{M^{2}} &=&\frac{8(b-2)(c-2)b}{(b+c)^{3}}+\frac{4(b-2)}{b+c}%
x+x^{2}-\frac{2b\sqrt{(b^{2}-4)(c^{2}-4)}\alpha \cos \theta }{(b+c)^{2}}%
+\alpha ^{2}\cos ^{2}\theta ,  \notag \\
\frac{H_{2}}{M^{2}} &=&\frac{8(b-2)(c-2)c}{(b+c)^{3}}+\frac{4(c-2)}{b+c}%
x+x^{2}-\frac{2c\sqrt{(b^{2}-4)(c^{2}-4)}\alpha \cos \theta }{(b+c)^{2}}%
+\alpha ^{2}\cos ^{2}\theta ,  \notag \\
\frac{H_{3}}{M^{2}} &=&x^{2}+\alpha ^{2}\cos ^{2}\theta -\frac{8x}{b+c},%
\text{ \ \ \ \ }\frac{\Delta }{M^{2}}=x^{2}+\alpha ^{2}-\frac{8x}{b+c},
\notag \\
\frac{H_{4}}{M^{3}} &=&\frac{2\sqrt{bc}\Big[(bc+4)(b+c)x-4(b-2)(c-2)\Big]%
\alpha \sin ^{2}\theta }{(b+c)^{3}},  \notag \\
m &=&\frac{4M}{b+c},\qquad \epsilon ^{2}=\frac{4c(c^{2}-4)}{(b+c)^{3}}%
,\qquad \mu ^{2}=\frac{4b(b^{2}-4)}{(b+c)^{3}},\qquad J=\frac{\sqrt{bc}(bc+4)%
}{(b+c)^{2}}M^{2}\alpha
\end{eqnarray}%
One of the advantages of this notation is a simple understanding of physical
properties of the solution. Indeed, the physical properties of this
spacetime can be explained in terms of $b$, $c$ and $\alpha $ parameters
more clearly and further comparisons with the Kerr-Newman BH can be made
possible. We should also note that some of the observational constraints on
free parameters and physical properties of the mentioned KK BH solution
(such as analysis of the gyroscope precession frequency \cite{KK9}, X-ray
reflection spectroscopy \cite{Xray}, and Shadow, quasinormal modes and
quasiperiodic oscillations \cite{Ghasemi}) have been studied before.


\subsection{Thermodynamics}\label{Thermodynamics}

Now, we turn to consider thermodynamics of the four dimensional KK BH. We
start with the calculation of horizon area ($\mathcal{A}$) using its
definition \cite{KK4}
\begin{equation*}
\mathcal{A}=\int_{0}^{\pi }d\theta \int_{0}^{2\pi }\sqrt{g_{_{\theta \theta
}}g_{_{\phi \phi }}}_{\big|_{r=r_{+}}}d\phi =-\frac{\pi \sqrt{bc}}{{\left(
b+c\right) ^{2}}}\,{\ }\left[ {{2+\frac{\left( bc+4\right) }{\left(
b+c\right) }\sqrt{1-\left( \frac{\left( b+c\right) {\alpha }}{2}\right) ^{2}}%
}}\right] M^{2}
\end{equation*}
In \cite{KK10}, the author proved that the entropy of KK BH obeys the area
formula which is given as \cite{KK4}
\begin{equation}
S=\frac{\mathcal{A}}{4\pi }=\frac{\pi {M}^{2}\sqrt{bc}}{2\left( b+c\right) }%
\left[ \frac{{x_{+}^{6}}-{\alpha }^{6}}{\left( {x_{+}^{2}}+{\alpha }%
^{2}\right) x_{+}^{2}}-\frac{2\left( {x_{+}^{2}}+{\alpha }^{2}\right) }{x_{+}%
}+{\alpha }^{2}+\frac{4\,bc}{\left( b+c\right) ^{2}}\right].  \label{Sr}
\end{equation}%
Here $x_{+}$ denotes the event horizon. If $x_+$ is large than
\begin{equation}
S\Big|_{x_{+}\rightarrow \infty }=\frac{\pi {M}^{2}\sqrt{bc}}{\left(
b+c\right) }\left[ \frac{{x}_{+}^{2}}{2}-x_{+}+{\frac{2\,bc}{\left(
b+c\right) ^{2}}}-{\frac{{\alpha }^{2}}{x_{+}}}\right] +O\left( \frac{1}{%
x_{+}^{2}}\right),  \label{largS}
\end{equation}
which confirms an expected result $S \propto r^2_{+}$ for large values of $%
r_{+}$ since this BH behaves similar to four dimensional static BH solutions.

The rotational velocity of the BH horizon ($\Omega =-g_{_{\phi t}}/g_{_{\phi
\phi }}$) reads
\begin{equation}
\Omega =\frac{\left( b+c\right) ^{2}\alpha }{2M\,\sqrt{bc}}\left[ 2+\frac{%
\left( bc+4\right) }{\left( b+c\right) }\sqrt{1-\left( \frac{\left(
b+c\right) {\alpha }}{2}\right) ^{2}}\right] ^{-1}
\end{equation}%
It is worthy to recall that $\Delta =0$ whenever we want to calculate $%
\Omega $.  One of the most important thermodynamic parameters in BH physics
is temperature. It is shown that \cite{KK10} temperature can be calculated
using the Euclidean method or surface gravity method, however, in our paper
we shall use the later one. The common Hawking temperature is related to
surface gravity of BHs ($T=\kappa /2\pi $). In our case, we encounter with a
stationary and axisymmetric solution, and corresponding to the Killing
vector field is $\chi =\partial /\partial t+\Omega \,\partial /\partial \phi
$ the surface gravity reads
\begin{equation}
\kappa =\sqrt{-\frac{1}{2}(\partial _{\mu }\chi _{\nu })(\partial ^{\mu
}\chi ^{\nu })}
\end{equation}%
It is easy to show that Hawking temperature is given by
\begin{equation}
T=\frac{\kappa }{4\pi }=\frac{x_{+}\left( {x_{+}^{4}}-{\alpha }^{4}\right)
\left( b+c\right) }{2M\pi \sqrt{bc}}\left[ \left( {\alpha }^{2}+{\frac{4bc}{%
\left( b+c\right) ^{2}}}\right) {x_{+}^{2}}\left( {x_{+}^{2}}-{\alpha }%
^{2}\right) +{x_{+}^{6}}-{\alpha }^{6}+2\,x_{+}\left( {x_{+}^{2}}+{\alpha }%
^{2}\right) ^{2}\right] ^{-1},
\end{equation}
which vanishes for $r_{+}=0$ or $r_{+}=a$. We should note that vanishing the
temperature after the origin may be related to the horizon of the extremal
BH.\newline
\begin{figure}[tbp]
{\includegraphics[scale=0.8]{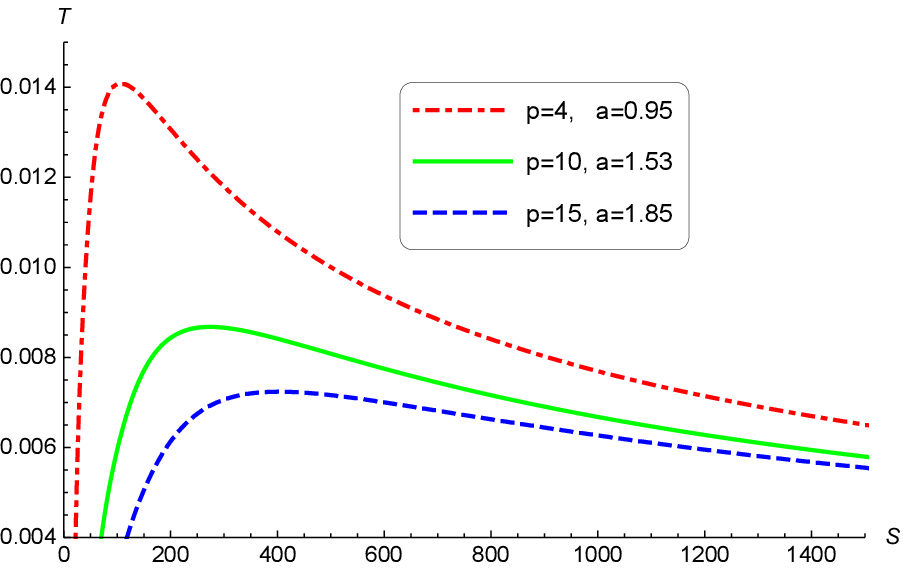}}\hspace*{.1cm} {%
\includegraphics[scale=0.8]{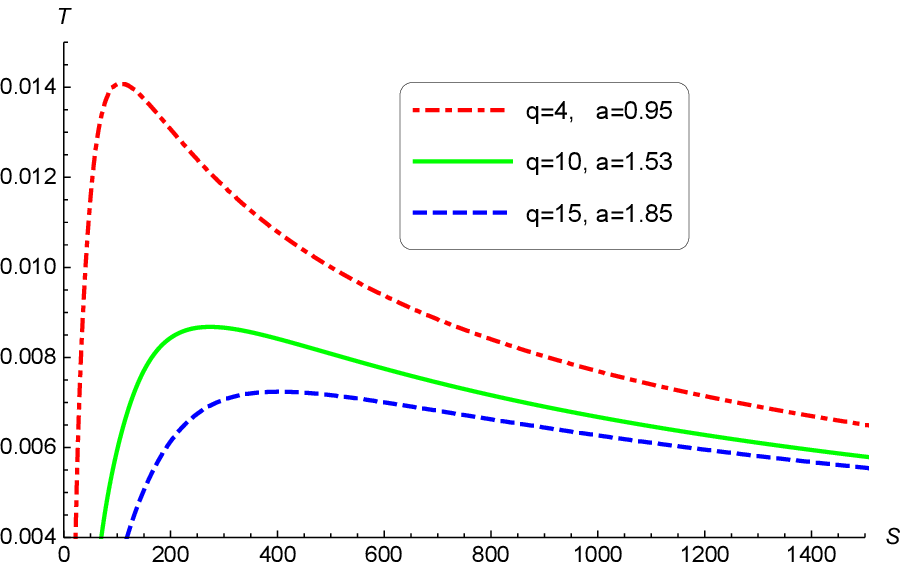}}
\caption{Behavior of temperature with respect to entropy for $q =
8$ (left) and $p=8$ (right)} \label{TS}
\end{figure}
According to Fig. \ref{TS} one finds the minimum of the entropy related with
the vanishing $T$ and it is consistent with the third law of thermodynamics.%
\newline
\begin{figure}[tbp]
\centering \subfigure[\; $q=5$ and $a=1.19$]  {%
\includegraphics[scale=0.9]{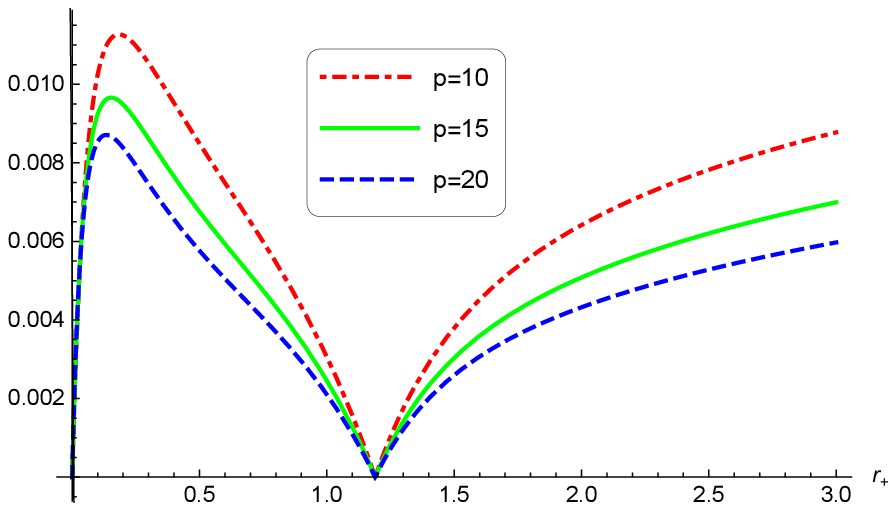}\label{temP}}\hspace*{.1cm}\subfigure[\;
$p=5$ and $a=1.19$]  {\includegraphics[scale=0.9]{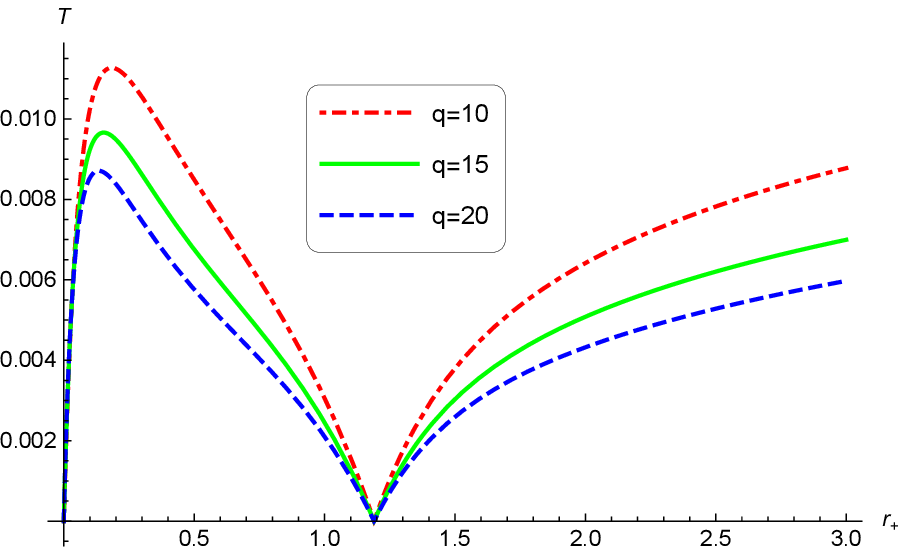}\label{temQ}} %
\subfigure[\; $p=4$ and $q=10$ ]  {\includegraphics[scale=1]{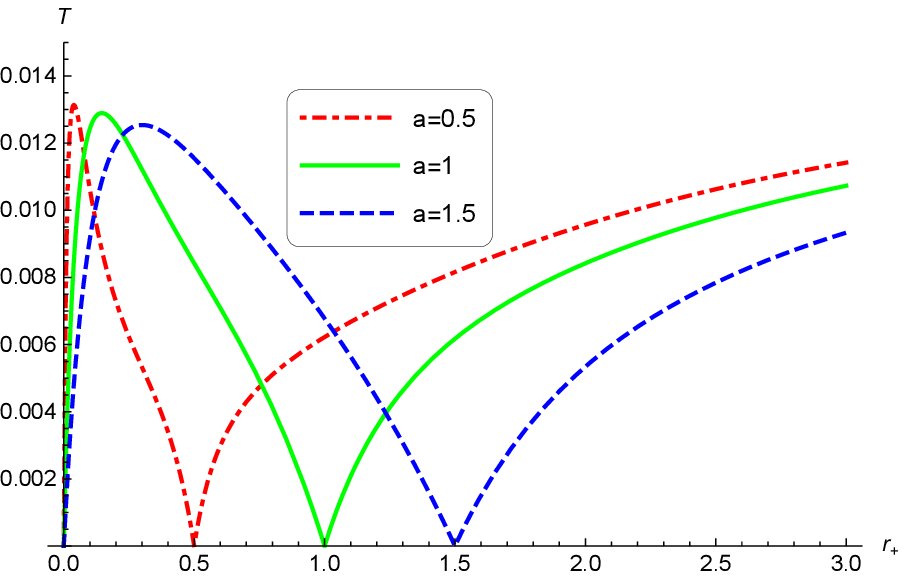}\label%
{temA}}\hspace*{.1cm}
\caption{Behavior of temperature with respect to $r_+$}
\label{tem}
\end{figure}
From Fig.\ref{tem}, it is clear that the position of the root and local
maximum of $T$ can change based on the metric parameters. For instance, one
finds that by increasing the spin parameter $a$, temperature vanishes for
larger $x_+$. It is worth noting that, changing in the magnetic (electric)
parameter does not alter the position of the root.

Besides, the expansion of temperature for large radius observes as
\begin{equation}
T\Big|_{x_{+}\rightarrow \infty }=\frac{\left( b+c\right) }{M\pi \sqrt{bc}}
\left( \frac{1}{2x_{+}}-\frac{1}{x_{+}^{2}}\right) +O\left( \frac{1}{
x_{+}^{3}}\right).  \label{largT}
\end{equation}
Interestingly enough, from (\ref{largS}) and (\ref{largT}) one can find that
for $x_{+}\to\infty$, the temperature vanishes but the entropy diverges! It
may offer the existence of infinite radius for KK BH \cite{KK10}.

The first law of BH thermodynamics can be verified as \cite{KK4}%
\begin{equation}
dM=TdS+UdQ+\Phi dP+\Omega dJ,  \label{1stLaw}
\end{equation}%
in which $U$ and $\Phi $ are electric and magnetic potential respectively,
given by
\begin{equation}
U=\left( \frac{\partial M}{\partial Q}\right) _{S,P,J},\text{ \ \ \ and \ \
\ }\Phi =\left( \frac{\partial M}{\partial P}\right) _{S,Q,J}.
\end{equation}%
After some manipulations, one can find the following explicit form of $U$\
and $\Phi $%
\begin{eqnarray}
&&U=\pi TM\,\sqrt{{\frac{\left( {c}^{2}-4\right) b}{\left( b+c\right) ^{3}}}}%
\left( b+{\frac{4}{\sqrt{4-{\alpha }^{2}\left( b+c\right) ^{2}}}}\right) , \\%
[10pt]
&&\Phi =\pi TM\,\sqrt{{\frac{\left( {b}^{2}-4\right) c}{\left( b+c\right)
^{3}}}}\left( c+{\frac{4}{\sqrt{4-{\alpha }^{2}\left( b+c\right) ^{2}}}}%
\right) .
\end{eqnarray}%
It is notable that for $p=2m$ ($q=2m$) which is equal to $b=2$ ($c=2$), the
magnetic (electric) potential vanishes. Also, both potential vanish when $%
x_{_{+}}\rightarrow \infty $ since%
\begin{eqnarray}
\Phi \Big|_{x_{_{+}}\rightarrow \infty } &=&\sqrt{{\frac{{b}^{2}-4}{\left(
b+c\right) b}}}\left( c+{\frac{4}{\sqrt{-{\alpha }^{2}\left( b+c\right)
^{2}+4}}}\right) \left( \frac{1}{2x_{_{+}}}-\frac{1}{x_{_{+}}^{2}}\right)
+O\left( \frac{1}{x_{_{+}}^{3}}\right)   \notag \\
U\Big|_{x_{_{+}}\rightarrow \infty } &=&\sqrt{{\frac{{c}^{2}-4}{\left(
b+c\right) c}}}\left( b+{\frac{4}{\sqrt{-{\alpha }^{2}\left( b+c\right)
^{2}+4}}}\right) \left( \frac{1}{2x_{_{+}}}-\frac{1}{x_{_{+}}^{2}}\right)
+O\left( \frac{1}{x_{_{+}}^{3}}\right) .
\end{eqnarray}%
which is expected for localized charged objects.

\section{Thermal stability via canonical ensemble approach}\label{Stability}

We are in a position to study thermal stability and phase transition of
solutions. By looking at the behavior of the heat capacity in the presence
of positive temperature, we predict criteria to have thermally stable BHs.
This approach is known as stability in canonical ensemble. When system is
unstable the phase transitions usually take place. In other words, unstable
systems go under a phase transition to acquire stability. Discontinuity of
heat capacity marks the second order phase transition in BHs \cite{hc1}.

The heat capacity for fixed values of extensive quantities obeys
\begin{equation}
C_{P,Q,J}=T\frac{\partial S}{\partial T}\Big|_{P,QJ}  \label{C}
\end{equation}%
In our case, entropy is not an explicit function of temperature, instead
they have common variables. Therefore we use chain rule of derivatives
to compute $C_{P,Q,J}$. It is notable that $P,$ $Q$, and $J$ are constants,
simultaneously, so we consider $dQ=dJ=dP=0$ to compute the heat capacity.
For the sake of complexity, we use some figures to analyze the treatment of
the heat capacity. It is important to note that we use different scales for
temperature to make it comparable with the heat capacity. In order to find
the critical point of heat capacity, we should use the first and second
derivative of temperature with respect to $r_{+}$. Firstly, we solve the
first derivative of temperature ($\frac{\partial T}{\partial x_{+}}=0$) to
obtain the critical angular momentum, then by substitution it in the second
derivative ($\frac{\partial ^{2}T}{\partial x_{+}^{2}}$=0), we may get the
critical horizon. These two functions are complicated and it is not a
trivial task to solve them, analytically. The practical solution is to use
the numerical method.

In table I, the critical values of spin parameter are presented. It is clear
that by increasing $p$ ($q$) when $q$ ($p$) is constant, the value of
critical $a$ increases. It is notable that replacing $p$ with $q$ does not
change the critical value of rotation parameter.
\begin{table}[tbp]
\caption{Table I: critical values of rotation parameter $a$}
\begin{center}
\begin{tabular}{|c|c|c|}
\hline
p & q & $a_{crit}$ \\ \hline
8 & 4 & 0.95 \\ \hline
9 & 4 & 1.00 \\ \hline
10 & 4 & 1.05 \\ \hline
20 & 4 & 1.39 \\ \hline
10 & 10 & 1.72 \\ \hline
\end{tabular}
\qquad
\begin{tabular}{|c|c|c|}
\hline
p & q & $a_{crit}$ \\ \hline
4 & 8 & 0.95 \\ \hline
4 & 9 & 1.00 \\ \hline
4 & 10 & 1.05 \\ \hline
4 & 20 & 1.39 \\ \hline
5 & 5 & 0.86 \\ \hline
\end{tabular}%
\end{center}
\end{table}

\begin{figure}[tbp]
\centering \subfigure[$p=4$, $q=8$ and $(a_{_{crit}}=0.95)$]
{\includegraphics[scale=0.9]{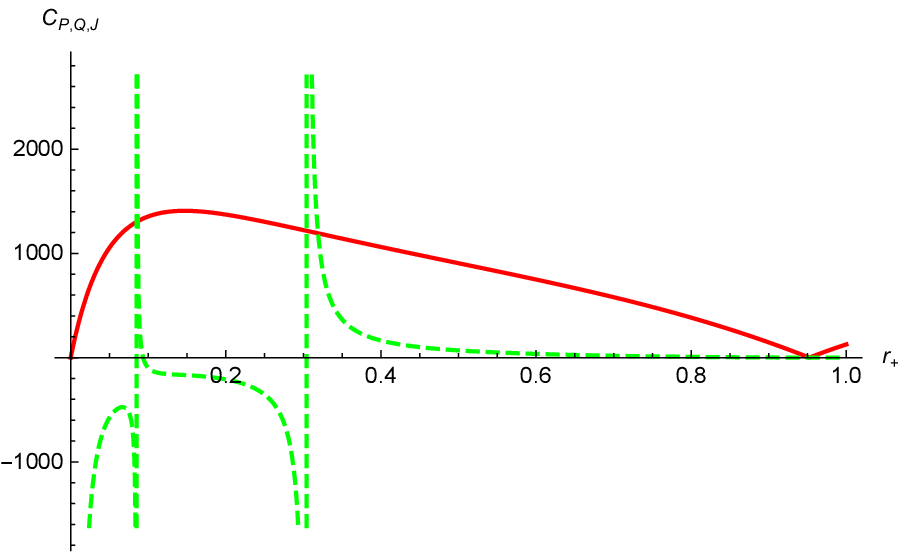}\label{crita}}\hspace*{.1cm}
\subfigure[$p=10$, $q=8$ and
$(a_{_{crit}}=1.53)$]{\includegraphics[scale=0.9]{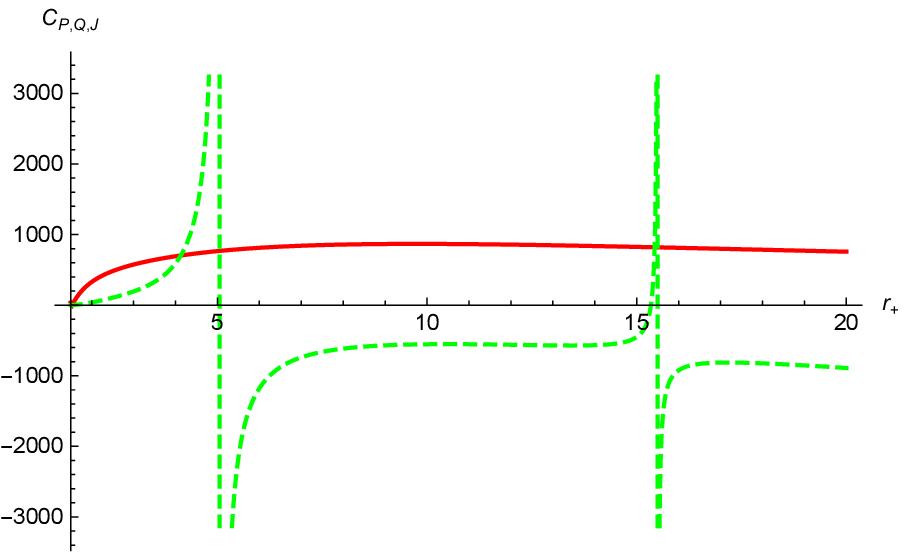}\label{critb}}
\subfigure[$p=15$, $q=8$ and $(a_{_{crit}}=1.85)$
]{\includegraphics[scale=1]{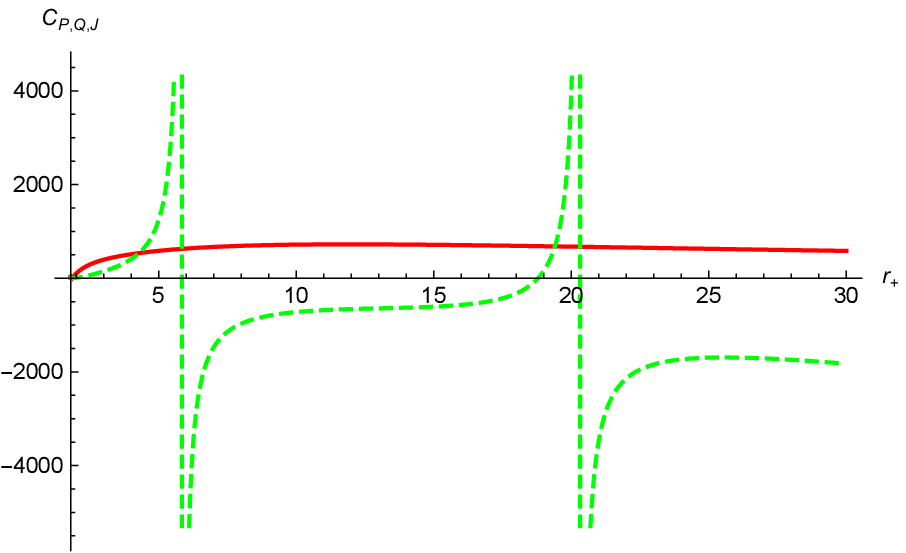}\label{crit1c}} \hspace*{.1cm}
\caption{Behavior of $C_{_{P,Q,J}}$ (green curve) and $10^{5}T$ (red curve)
with respect to $r_+$}
\label{crit}
\end{figure}

Regarding Eq. (\ref{C}), one may find positive heat capacity for negative $T$
and $\frac{\partial S}{\partial T}\Big|_{P,QJ}<0$ which is not physical
stability. In order to remove such an ambiguity, we plot both temperature
and heat capacity in Fig. \ref{crit}. By adjusting the electric and magnetic
charge parameters, we can find the critical rotation parameter $a_{crit}$.
Numerical calculations show that increasing in the magnetic (electric)
charge makes the critical rotation parameters get larger. By exchanging the
value of two parameters no change in $a_{crit}$ is observed.

Figure \ref{crit} shows two divergencies and one zero value in the heat
capacity function in the presence of positive temperature, which their
positions change by increasing in the magnetic (electric) parameter. As we
mentioned before, the only acceptable $C_x$ ($x$ means $P,Q,J)$ is positive
one, so the stable BH is only allowed to have limited radii. Based on the
figures, the heat capacity is negative after the final divergency, it
results that large BHs are not stable thermodynamically.

Another interesting note is related to the position of two divergencies to
each other. Increasing in the metric parameters makes them farther however,
it does not have much effect on the allowable values of the radius. The
final point is that as the divergencies occur between positive and negative
values of $C_x$ so the plots do not predict first or second order phase
transition but the Davis phase transition is possible.

Finally, we plot heat capacity when magnetic charge parameter is fixed but
the curves do not contain additional information.


\section{Quasinormal modes}\label{QNMs}

\subsection{Setup}

Here, we consider a massive scalar perturbation in the background geometry
of $4$-dimensional static KK BHs and obtain the QN frequencies by employing
the WKB approximation \cite{Schutz,IyerWill,Konoplya6th,Matyjasek13th}. The
line element of $4$-dimensional KK BHs (\ref{Metric4d}) for the static case $%
a=0$\ reduces to%
\begin{equation}
ds_{4}^{2}=-f(r)dt^{2}+\frac{dr^{2}}{f(r)}+g(r)\left( d\theta ^{2}+\sin
\theta d\phi ^{2}\right) ,  \label{StaticMetric}
\end{equation}%
with $f(r)=H_{3}/g(r)$ and $g(r)=\sqrt{H_{1}H_{2}}$. The equation of motion
for a minimally coupled massive scalar field $\Psi $\ is given by the
following Klein--Gordon equation
\begin{equation}
\square \Psi -\mu ^{2}\Psi =0,  \label{KGeq}
\end{equation}%
in which $\mu $ is the mass of the scalar field $\Psi $ and $\square =\nabla
_{\nu }\nabla ^{\nu }$. It is notable to mention that we cannot obtain a
second-order Schr\"{o}dinger-like wave equation for the radial part of
perturbations by expanding the scalar field versus either spherical or
spheroidal harmonics. In order to find a Schr\"{o}dinger-like master
equation, hence being able to use the WKB approximation, we first define $R=%
\sqrt{g(r)}$\ and rewrite the metric (\ref{StaticMetric})\ in the following
form
\begin{equation}
ds_{4}^{2}=-f(R)dt^{2}+\frac{dR^{2}}{f(R)h^{2}(R)}+R^{2}\left( d\theta
^{2}+\sin \theta d\phi ^{2}\right) ,  \label{SSmetric}
\end{equation}%
where $f(R)$\ and $h(R)$\ can be obtained by converting $r$\ versus $R$\
through $R=\sqrt{g(r)}$. However, we should note that to calculate $r(R)$,
the equation $R-\sqrt{g(r)}=0$\ has four independent solutions. Here, we
choose the solution which maps the event horizon $r_{+}$ of (\ref%
{StaticMetric}) to a positive definite event horizon $R_{+}$ for (\ref%
{SSmetric}).

Now, by expanding the scalar field eigenfunction $\Psi $ in the form%
\begin{equation}
\Psi \left( t,R,\theta ,\varphi \right) =\sum_{l,m}\frac{1}{R}\psi
_{l}\left( R\right) Y_{l,m}\left( \theta ,\varphi \right) e^{-i\omega t},
\label{expansion}
\end{equation}%
which $Y_{l,m}\left( \theta ,\varphi \right) $ denotes the spherical
harmonics on $S^{2}$, we can find that the equation of motion (\ref{KGeq})
reduces to a wavelike equation for the radial part $\psi _{l}\left( R\right)
$ as follows
\begin{equation}
\left[ \partial _{R_{\ast }}^{2}+\omega ^{2}-V_{l}\left( R\right) \right]
\psi _{l}\left( R_{\ast }\right) =0.  \label{Weq}
\end{equation}

In this equation, $R_{\ast }$ is the tortoise coordinate
\begin{equation}
R_{\ast }=\int \frac{dR}{f(R)h(R)},  \label{tortoise}
\end{equation}%
and the effective potential $V_{l}\left( R\right) $ is given by
\begin{equation}
V_{l}\left( R\right) =f\left( R\right) \left[ \mu ^{2}+\frac{l\left(
l+1\right) }{R^{2}}+\frac{h\left( R\right) }{R}\partial _{R}\left[ f(R)h(R)%
\right] \right] ,  \label{EP}
\end{equation}%
where $l$ is the multipole number.

By imposing some proper boundary conditions on the master wave equation (\ref%
{Weq}), we can find a discrete set of eigenvalues $\omega $. The quasinormal
boundary conditions imply that the wave at the event horizon is purely
incoming and the modes are purely outgoing at spacial infinity.\
\begin{equation}
\begin{array}{c}
\psi _{l}\left( R\right) \sim e^{-i\omega R_{\ast }}\ \ \ \ \ \ as\ \ \ \ \
\ R_{\ast }\rightarrow -\infty \ (r\rightarrow r_{+}) \\
\psi _{l}\left( R\right) \sim e^{i\omega R_{\ast }}\ \ \ \ \ \ \ as\ \ \ \ \
\ R_{\ast }\rightarrow \infty \ (r\rightarrow \infty )%
\end{array}%
,  \label{bc}
\end{equation}%
and we should consider these boundary conditions to obtain the QNMs spectrum.

\begin{figure}[tbp]
$%
\begin{array}{ccc}
\epsfxsize=7.5cm \epsffile{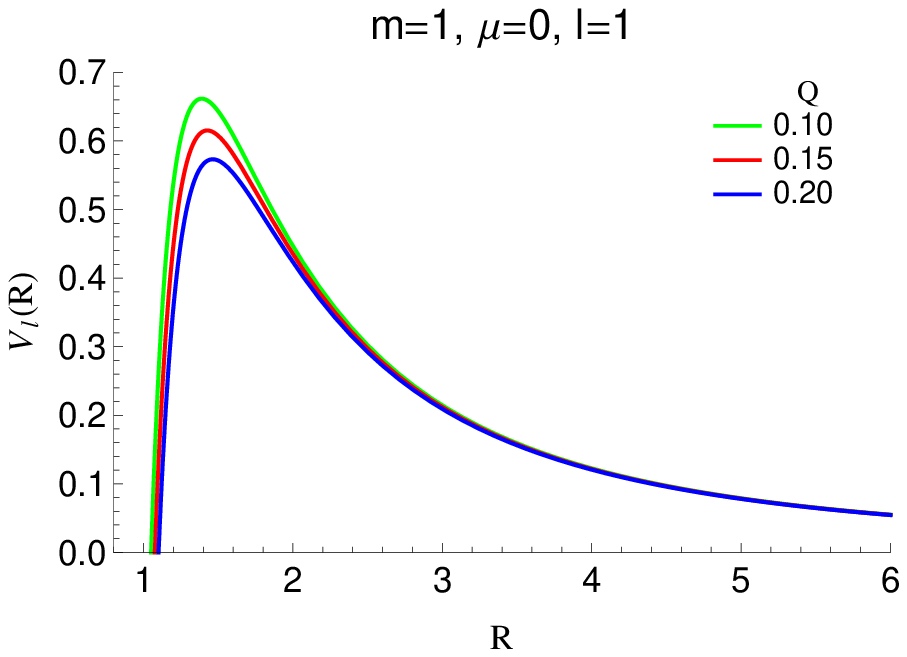} & \epsfxsize=7.5cm \epsffile{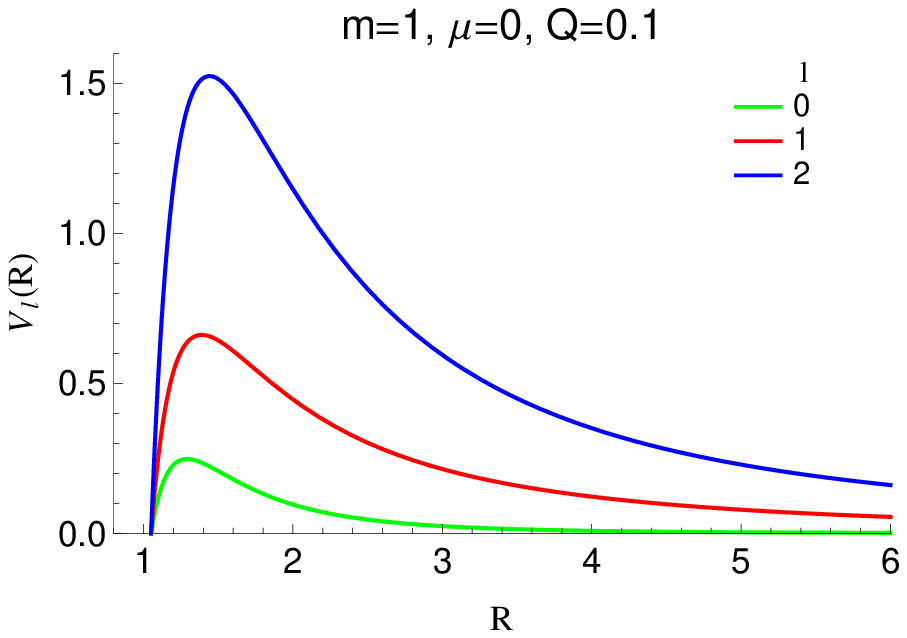} &
\end{array}
$%
\caption{Profiles of the effective potential versus the radial coordinate.
The potential forms a barrier and vanishes at both infinities.}
\label{Pot}
\end{figure}

\subsection{WKB approximation}

In this paper, we use the WKB approximation to calculate the QN modes. This
approximation is based on the matching of WKB expansion of the modes $\psi
_{l}\left( r_{\ast }\right) $ at the event horizon and spatial infinity with
the Taylor expansion of the effective potential (\ref{EP}) near the peak of
the potential barrier through two turning points at $\omega ^{2}-V_{l}\left(
R\right) =0$. Thus, we can use the WKB approximation to calculate the QN
frequencies for potentials that form a potential barrier and takes constant
and/or zero values at the event horizon and spatial infinity. The WKB
approximation was first applied to the problem of scattering around black
holes \cite{Schutz}, and subsequently extended to the third-order \cite%
{IyerWill}, $6$th order \cite{Konoplya6th} and $13$th order \cite%
{Matyjasek13th}. The $13$th order of WKB approximation is given by the
following formula%
\begin{equation}
\omega ^{2}=V_{0}+\sum_{j=1}^{6}\Omega _{2j}-i\sqrt{-2V_{0}^{\prime \prime }}%
\left( n+\frac{1}{2}\right) \left( 1+\sum_{j=1}^{6}\Omega _{2j+1}\right) ;\
\ \ \ \ \ n=0,1,2,...,  \label{WKB}
\end{equation}%
where $V_{0}$\ is the maximum value of the effective potential, $\Omega _{j}$%
's are the WKB correction terms of the $j$th order, and $n$\ is the overtone
number. It is worthwhile to mention that the WKB formula does not give
reliable frequencies for $n\geq l$, whereas it leads to quite accurate
values for $n<l$\ \ and exact values in the eikonal limit $l\rightarrow
\infty $. We use this formula up to the $13$th order to calculate the QN
frequencies of perturbations.

However, at the first step, we should note that the relation $r(R)$\ is
generally quite complicated and leads to a cumbersome form for the effective
potential. Thus, obtaining the QN frequencies even by using the third-order
WKB formula is time-consuming. But, fortunately, for an equal value of $p$\
and $q$ ($p=Q=q$), we receive a quite simple relation as $r(R)=m+R-Q/2$\ and
the effective potential takes a more simple form as follows%
\begin{equation}
V_{l}\left( R\right) =f(R)\left( \mu ^{2}+\frac{l\left( l+1\right) }{R^{2}}+%
\frac{f^{\prime }(R)}{R}\right) ,  \label{SimpleP}
\end{equation}%
with%
\begin{equation}
f(R)=-\frac{m^{2}}{R^{2}}+\frac{\left( Q-2R\right) ^{2}}{4R^{2}}.
\label{SECMetric}
\end{equation}

We shall use this potential to calculate the QNMs. Figure
\ref{Pot} shows the behavior of this effective potential
(\ref{SimpleP}) versus radial coordinate for different values of
charge $Q$ and the multipole number $l$. The potential forms a
barrier and vanishes at the event horizon and spatial infinity,
thus we can use the WKB formula to calculate the QN frequencies.

As we have mentioned before, the WKB formula usually gives the best accuracy
for $l>n$ and it provides an accurate and economic way to compute the QN
frequencies \cite{Konoplya6th,wkbError}. In this regards we compare two
sequential orders of the formula (\ref{WKB}) to estimate the error of the
WKB approximation. However, since each WKB correction term affects either
the real or imaginary part of the squared frequencies, we should use the
following quantity \cite{wkbError}%
\begin{equation}
\Delta _{k}=\frac{\left\vert \omega _{k+1}-\omega _{k-1}\right\vert }{2},
\label{error}
\end{equation}%
to obtain the error estimation of $\omega _{k}$ that is calculated with the
WKB formula of the order $k$, and $\Delta _{k}$\ gives the WKB order in
which the error is minimal. Therefore, we can use the error estimation (\ref%
{error}) to find the WKB order which gives the most accurate approximation
for the QN modes.

In table $II$, we show the QN frequencies and the error estimation
of the WKB formula for the fundamental QN modes. From this table,
we see that the best
order of the WKB formula for calculating the QN frequency for $Q=0.1$ is $7$%
th-order whereas the QN frequency for $Q=0.2$ has the best accuracy with the
help of the $5$th-order. Thus, the minimum error of the WKB formula depends
on the charge value $Q$. The oscillations increase and the modes live longer
as the charge $Q$\ decreases.

\begin{center}
\begin{tabular}{|cccc|c|c|c|}
\hline\hline
$k$ & \multicolumn{1}{|c}{$\omega _{k}$} & \multicolumn{1}{|c}{$\Delta _{k}$
$\times 10^{-3}$} & \multicolumn{1}{|c|}{} & $k$ & $\omega _{k}$ & $\Delta
_{k}$ $\times 10^{-3}$ \\ \hline\hline
$2$ & \multicolumn{1}{r}{$0.7156-0.4043i$} & $99.6$ & \multicolumn{1}{r|}{}
& \multicolumn{1}{r}{$2$} & \multicolumn{1}{r}{$0.6725-0.4228i$} & $99.3$ \\
\hline
$3$ & \multicolumn{1}{r}{$0.6786-0.3344i$} & $41.2$ & \multicolumn{1}{r|}{}
& \multicolumn{1}{r}{$3$} & \multicolumn{1}{r}{$0.6328-0.3563i$} & $40.7$ \\
\hline
$4$ & \multicolumn{1}{r}{$0.7031-0.3228i$} & $14.7$ & \multicolumn{1}{r|}{}
& \multicolumn{1}{r}{$4$} & \multicolumn{1}{r}{$0.6577-0.3428i$} & $15.6$ \\
\hline
$5$ & \multicolumn{1}{r}{$0.7079-0.3331i$} & $7.43$ & \multicolumn{1}{r|}{}
& \multicolumn{1}{r}{$5$} & \multicolumn{1}{r}{$\mathbf{0.6639-0.3547}i$} & $%
\mathbf{9.11}$ \\ \hline
$6$ & \multicolumn{1}{r}{$0.6994-0.3372i$} & $6.49$ & \multicolumn{1}{r|}{}
& \multicolumn{1}{r}{$6$} & \multicolumn{1}{r}{$0.6532-0.3605i$} & $9.52$ \\
\hline
$7$ & \multicolumn{1}{r}{$\mathbf{0.6955-0.3291}i$} & $\mathbf{4.96}$ &
\multicolumn{1}{r|}{} & \multicolumn{1}{r}{$7$} & \multicolumn{1}{r}{$%
0.6462-0.3477i$} & $11.4$ \\ \hline
$8$ & \multicolumn{1}{r}{$0.6995-0.3273i$} & $6.76$ & \multicolumn{1}{r|}{}
& \multicolumn{1}{r}{$8$} & \multicolumn{1}{r}{$0.6618-0.3394i$} & $12.8$ \\
\hline
$9$ & \multicolumn{1}{r}{$0.6942-0.3157i$} & $34.0$ & \multicolumn{1}{r|}{}
& \multicolumn{1}{r}{$9$} & \multicolumn{1}{r}{$0.6705-0.3560i$} & $9.95$ \\
\hline
$10$ & \multicolumn{1}{r}{$0.7562-0.2898i$} & $112$ & \multicolumn{1}{r|}{}
& \multicolumn{1}{r}{$10$} & \multicolumn{1}{r}{$0.6646-0.3592i$} & $32.9$
\\ \hline
$11$ & \multicolumn{1}{r}{$0.8470-0.4792i$} & $360$ & \multicolumn{1}{r|}{}
& \multicolumn{1}{r}{$11$} & \multicolumn{1}{r}{$0.6971-0.4162i$} & $205$ \\
\hline
$12$ & \multicolumn{1}{r}{$0.4344-0.9342i$} & $891$ & \multicolumn{1}{r|}{}
& \multicolumn{1}{r}{$12$} & \multicolumn{1}{r}{$0.4209-0.6893i$} & $767$ \\
\hline\hline
\end{tabular}%
\vspace{0.2cm}

Table $II$: The fundamental modes calculated by the WKB formula of
different orders for $m=1$, $\mu =0$, $n=0$, $l=1$,\ and $Q=0.1$
left ($Q=0.2$ right). The minimal error estimation is given in
bold form.
\end{center}

\subsection{Anomalous decay rate of QN modes}

One of the motivations for considering the test massive fields comes from
the fact that depending on the mass of the scalar field, the QNMs either
grow or decay with increasing multipole number $l$. This novel behavior
first was uncovered for Schwarzschild BH \cite{ADR}, and then confirmed for
the Reissner-Nordstr\"{o}m BH \cite{ADRrn}, Schwarzschild-dS\ spacetime, and
BH solutions\ in conformal Weyl gravity \cite{WeylGravityQRM}. This
anomalous behavior is due to the presence of a sub-leading $\mu ^{2}$-term
in the eikonal expression of $\omega _{i}$ \cite{ADR}. In this scenario,
there is a critical scalar mass $\tilde{\mu}$\ such that $\omega _{i}$
increases (decreases) with increasing in $l$ for $\mu >\tilde{\mu}$\ ($\mu <%
\tilde{\mu}$). For low-$l$ values, the critical mass $\tilde{\mu}$ decreases
when $l$\ increases, but there is a fixed critical mass for large-$l$ values.

\begin{figure}[tbp]
$%
\begin{array}{ccc}
\epsfxsize=7.5cm \epsffile{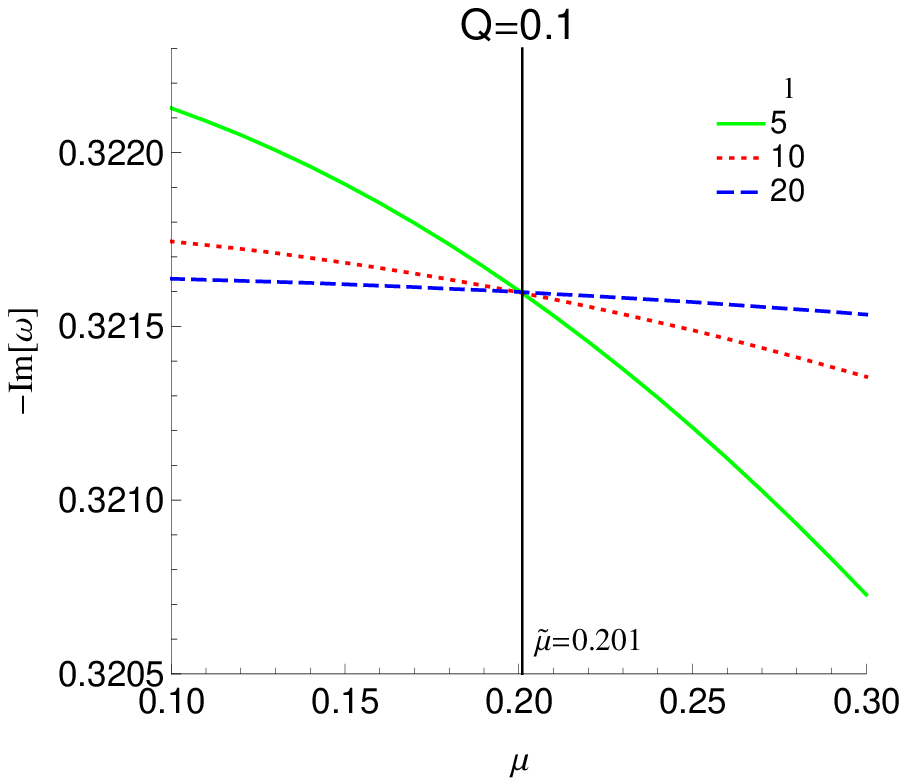} & \epsfxsize=7.5cm \epsffile{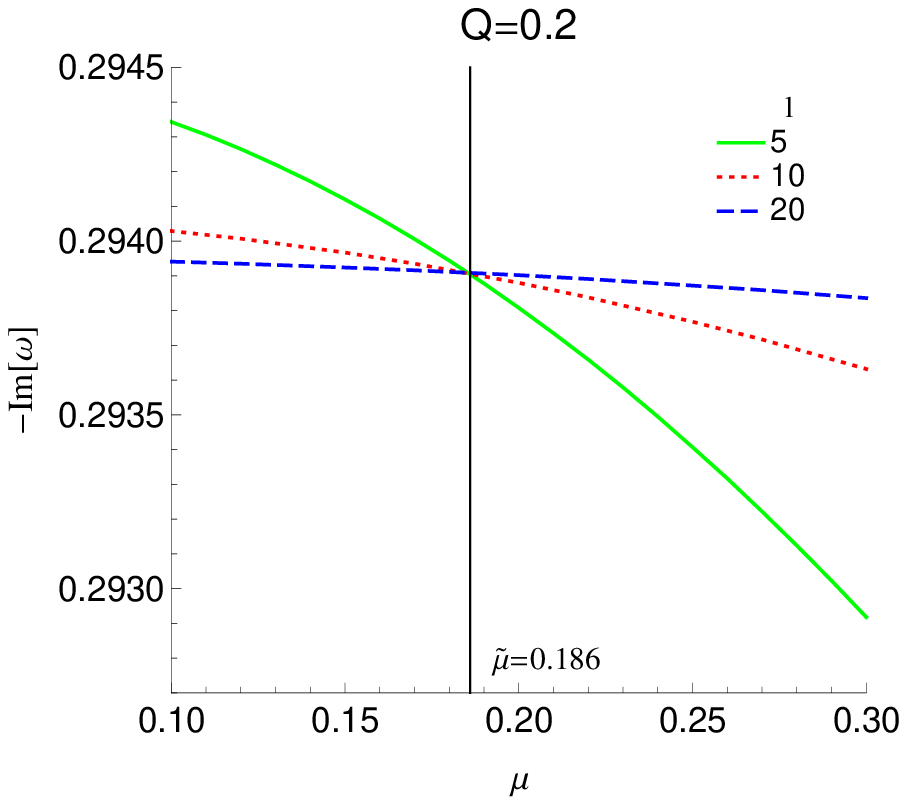}
&
\end{array}
$%
\caption{The imaginary part of the fundamental overtone versus $\protect\mu $
calculated by using the sixth order WKB formula for $m=1$. The vertical
black line indicates the critical mass $\tilde{\protect\mu}$ where the
curves cross each other for large-$l$ values.}
\label{Anomalous}
\end{figure}

Here, we numerically\ investigate the possibility of this anomalous behavior
for our BH case study with the line element (\ref{SSmetric}). Note that
since we are going to calculate the fundamental QN frequencies for large-$l$
values, the WKB approximation will lead to accurate results. Thus, we have
used the sixth-order WKB formula to plot the figure \ref{Anomalous}. This
figure shows the imaginary part of the QN modes $\omega _{i}$ as a function
of $\mu $ for different values of $l$\ and $Q$. From both panels, we find
that the curves cross over at a special mass $\tilde{\mu}$, and thus the QNM
spectrum of KK BHs in asymptotically flat spacetime contains this anomaly.
It is worthwhile to note that the charge parameter $Q$\ affects the critical
mass and $\tilde{\mu}$\ increases with decreasing in $Q$.

\subsection{Quasi-resonance modes}

\begin{figure}[tbp]
$%
\begin{array}{ccc}
\epsfxsize=7.5cm \epsffile{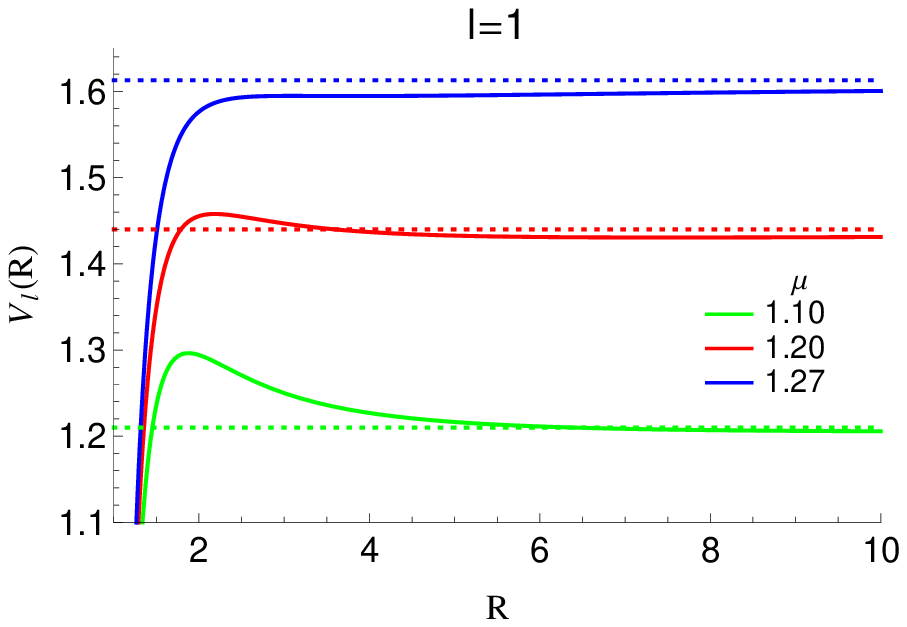} & \epsfxsize=7.5cm \epsffile{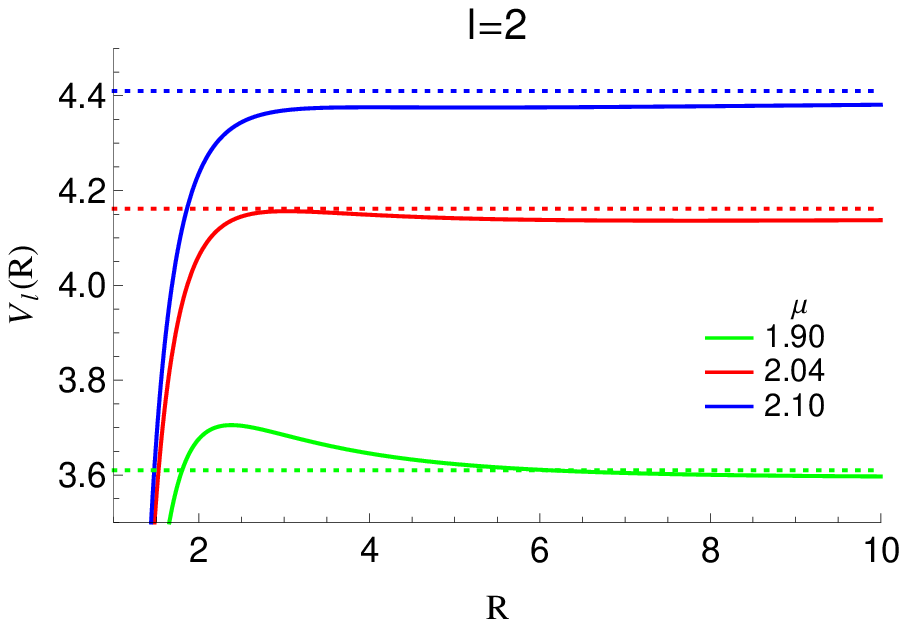}
&
\end{array}
$%
\caption{The effective potential for $m=1$ and $Q=0.1$. The potential takes
a constant value at the spacial infinity and the dotted horizontal lines
show these asymptotic values.}
\label{QRMsPot}
\end{figure}

In addition to the anomalous decay rate of QNMs related to massive test
fields, observing arbitrarily long life (purely real) modes is also one of
the interesting motivations for studying massive scalar fields \cite{RN-QRM}%
. These kinds of modes with vanishing imaginary parts are called
quasi-resonance modes. The oscillations do not decay in the quasi-resonances
and the situation is similar to the standing waves on a string. The
quasi-resonance modes were investigated for Schwarzschild BH \cite%
{SchwQRMs,higherSchwQRM}, Reissner-Nordstr\"{o}m BH \cite{Hod-rnQRM},
magnetized Schwarzschild BH \cite{magnetizedSchw-QRM}, Kerr geometry \cite%
{KerrQRM}, BHs in Einstein-Weyl gravity \cite{EinWeylGravityQRM}, and
wormholes \cite{WormQRM}. However, it is not possible to find these modes
for asymptotically dS spacetimes \cite{WeylGravityQRM,SchwDS-QRMs}. The
quasi-resonance modes can be found for special values of the field mass
whenever the effective potential is non-zero at the event horizon or spacial
infinity (see Fig. \ref{QRMsPot}\ for the profile of the effective potential
(\ref{SimpleP})\ of our BH case study). In this scenario, the QNMs disappear
and this happens just for lower overtones.

We recall that the WKB approximation provides quite a simple, powerful, and
accurate tool for studying the dynamical properties of BHs, such as the
scattering problems and QN modes for low overtones and high multipole
numbers. But, this method cannot be used for the calculation of
quasi-resonance modes in general and it just allows one to calculate large-$%
l $ QN frequencies of massive test fields close to the quasi-resonance
regime \cite{wkbError}. The reason is that the effective potential does not
have a local maximum for large values of the field mass, thus the WKB
expansion cannot be performed. However, as long as the asymptotic value of
the effective potential is lower than its peak, $\mu ^{2}<V_{0}$ (like the
green and red curves in Fig. \ref{QRMsPot}), the ordinary WKB formula (\ref%
{WKB})\ is accurate enough for $l\geq 1$, hence the error is neglectable
\cite{wkbError}.

In order to calculate the quasi-resonances by employing the WKB
approximation, we use an approach based on averaging of Pad\'{e}
approximations \cite{Matyjasek13th} which is developed for
quasi-resonances of the Schwarzschild BH \cite{wkbError} which can
considerably improve the accuracy of the quasi-resonance modes for
$\mu ^{2}>V_{0}$\ when the maximum of the potential still exists
(see the blue curves in Fig. \ref{QRMsPot}). The fundamental QN
modes calculated by averaging results obtained by Pad\'{e}
approximations of various orders and related standard deviation
(SD) formula are shown in tables $III$ and $IV$ for $l=1,2$ in
order. As we observe from these tables, the minimal SD changes
based on the scalar mass so that the higher orders lead to minimal
SD for low-$\mu $ values and the lower orders lead to minimal SD
for high-$\mu $\ values, unlike the Schwarzschild case in which
the SD formula of averaging Pad\'{e} approximates of $13$th order
is
minimal for all values of field mass (see table $XI$ of Ref. \cite{wkbError}%
).

\begin{center}
\begin{tabular}{|c|c|c|c|}
\hline\hline
$k $ & $\mu =1$ (SD$\times 10^{-5}$) & $\mu =1.2$ (SD$\times 10^{-4}$) & $%
\mu =1.27$ (SD$\times 10^{-4}$) \\ \hline\hline
$1$ & $1.0665-0.1624i~(2501\%)$ & $1.2054-0.0699i~(40.6\%)$ & $\mathbf{%
1.2628-0.0138}i~\mathbf{(1.50\%)}$ \\ \hline
$2$ & $0.9867-0.1181i~(1563\%)$ & $1.1397-0.0357i~(414\%)$ & $%
1.1389-0.0036i~(1235\%)$ \\ \hline
$3$ & $0.9743-0.1308i~(4.18\%)$ & $1.1117-0.0236i~(49.3\%)$ & $%
1.2624-0.0228i~(4.12\%)$ \\ \hline
$4$ & $0.9693-0.1276i~(302\%)$ & $1.1171-0.0311i~(21.8\%)$ & $%
1.1552-0.0071i~(1095\%)$ \\ \hline
$5$ & $0.9679-0.1325i~(132\%)$ & $1.1270-0.0168i~(130\%)$ & $%
1.0459-0.1030i~(3922\%)$ \\ \hline
$6$ & $0.9633-0.1254i~(159\%)$ & $\mathbf{1.1153-0.0273}i~\mathbf{(9.80\%)}$
& $1.1734-0.0076i~(892\%)$ \\ \hline
$7$ & $0.9644-0.1262i~(51.0\%)$ & $1.1095-0.0213i~(86.2\%)$ & $%
1.2153-0.3088i~(5496\%)$ \\ \hline
$8$ & $0.9638-0.1265i~(75.2\%)$ & $1.1105-0.0203i~(16.2\%)$ & $%
3.5935-0.0637i~(>10^{4}\%)$ \\ \hline
$9$ & $0.9636-0.1258i~(15.4\%)$ & $1.1110-0.0190i~(16.9\%)$ & $%
1.1962-0.2423i~(3483\%)$ \\ \hline
$10$ & $0.9631-0.1252i~(87.3\%)$ & $1.1109-0.0185i~(45.0\%)$ & $%
2.0596-0.0650i~(>10^{4}\%)$ \\ \hline
$11$ & $0.9637-0.1257i~(31.4\%)$ & $0.8534-1.5287i~(>10^{4}\%)$ & $%
1.0131-0.1821i~(4401\%)$ \\ \hline
$12$ & $0.9638-0.1258i~(13.0\%)$ & $1.1087-0.0169i~(39.8\%)$ & $%
1.7072-0.0701i~(2356\%)$ \\ \hline
$13$ & $\mathbf{0.9637-0.1258}i~\mathbf{(1.96\%)}$ & $1.1080-0.0194i~(24.5%
\%) $ & $5.4694+4.129i~(>10^{4}\%)$ \\ \hline\hline
\end{tabular}%
\vspace{0.2cm}

Table $II$: The fundamental modes calculated by averaging of Pad\'{e}
approximations for $m=1$, $Q=0.1$,\ and $l=1$. The minimal standard
deviation formula is given in bold.

\begin{tabular}{|cccc}
\hline\hline
$k$ & \multicolumn{1}{|c}{$\mu =1.8$ (SD$\times 10^{-6}$)} &
\multicolumn{1}{|c}{$\mu =1.9$ (SD$\times 10^{-5}$)} & \multicolumn{1}{|c|}{$%
\mu =2.1$ (SD$\times 10^{-5}$)} \\ \hline\hline
$1$ & \multicolumn{1}{r}{$1.8480-0.0934i~(4728\%)$} & \multicolumn{1}{r}{$%
1.9240-0.0681i~(241\%)$} & \multicolumn{1}{r}{$\mathbf{2.0918-0.0085}i~%
\mathbf{(3.49\%)}$} \\ \hline
$2$ & \multicolumn{1}{r}{$1.8038-0.0781i~(>10^{4}\%)$} & \multicolumn{1}{r}{$%
1.8845-0.0538i~(1240\%)$} & \multicolumn{1}{r}{$2.0432-0.0037i~(4808\%)$} \\
\hline
$3$ & \multicolumn{1}{r}{$1.7991-0.0814i~(295\%)$} & \multicolumn{1}{r}{$%
1.8781-0.0559i~(52.6\%)$} & \multicolumn{1}{r}{$2.0914-0.0145i~(10.0\%)$} \\
\hline
$4$ & \multicolumn{1}{r}{$1.7973-0.0809i~(496\%)$} & \multicolumn{1}{r}{$%
1.8766-0.0555i~(35.6\%)$} & \multicolumn{1}{r}{$2.0572-0.0046i~(3325\%)$} \\
\hline
$5$ & \multicolumn{1}{r}{$1.7974-0.0803i~(31.3\%)$} & \multicolumn{1}{r}{$%
1.8769-0.0547i~(17.6\%)$} & \multicolumn{1}{r}{$2.0358-0.1179i~(>10^{4}\%)$}
\\ \hline
$6$ & \multicolumn{1}{r}{$1.7975-0.0805i~(60.8\mathbf{\%)}$} &
\multicolumn{1}{r}{$1.8769-0.0546i~(5.04\mathbf{\%)}$} & \multicolumn{1}{r}{$%
2.0622-0.0052i~(2714\%)$} \\ \hline
$7$ & \multicolumn{1}{r}{$1.7974-0.0804i~(28.3\%)$} & \multicolumn{1}{r}{$%
1.8769-0.0547i~(14.1\%)$} & \multicolumn{1}{r}{$2.0613-0.0931i~(>10^{4}\%)$}
\\ \hline
$8$ & \multicolumn{1}{r}{$1.7974-0.0805i~(16.1\%)$} & \multicolumn{1}{r}{$%
1.8767-0.0550i~(13.2\%)$} & \multicolumn{1}{r}{$2.6307-0.0415i~(>10^{4}\%)$}
\\ \hline
$9$ & \multicolumn{1}{r}{$1.7973-0.0805i~(10.4\%)$} & \multicolumn{1}{r}{$%
\mathbf{1.8768-0.0550}i~\mathbf{(1.10\%)}$} & \multicolumn{1}{r}{$%
2.0617-0.0650i~(5125\%)$} \\ \hline
$10$ & \multicolumn{1}{r}{$1.7972-0.0804i~(186\%)$} & \multicolumn{1}{r}{$%
1.8767-0.0551i~(15.3\%)$} & \multicolumn{1}{r}{$1.8090-0.2306i~(>10^{4}\%)$}
\\ \hline
$11$ & \multicolumn{1}{r}{$1.7974-0.0805i~(7.97\%)$} & \multicolumn{1}{r}{$%
1.8767-0.0549i~(4.65\%)$} & \multicolumn{1}{r}{$3.7053+1.6704i~(>10^{4}\%)$}
\\ \hline
$12$ & \multicolumn{1}{r}{$\mathbf{1.7974-0.0805}i~\mathbf{(3.43\%)}$} &
\multicolumn{1}{r}{$1.8767-0.0548i~(17.7\%)$} & \multicolumn{1}{r}{$%
1.8096-0.4064i~(>10^{4}\%)$} \\ \hline
$13$ & \multicolumn{1}{r}{$1.7974-0.0805i~(5.74\%)$} & \multicolumn{1}{r}{$%
1.8767-0.0549i~(2.93\%)$} & \multicolumn{1}{r}{$2.7043+0.8904i~(>10^{4}\%)$}
\\ \hline\hline
\end{tabular}%
\vspace{0.2cm}

Table $III$: The fundamental modes calculated by averaging of Pad\'{e}
approximations for $m=1$, $Q=0.1$,\ and $l=2$. The minimal standard
deviation formula is given in bold.
\end{center}

\begin{figure}[tbp]
$%
\begin{array}{ccc}
\epsfxsize=7.5cm \epsffile{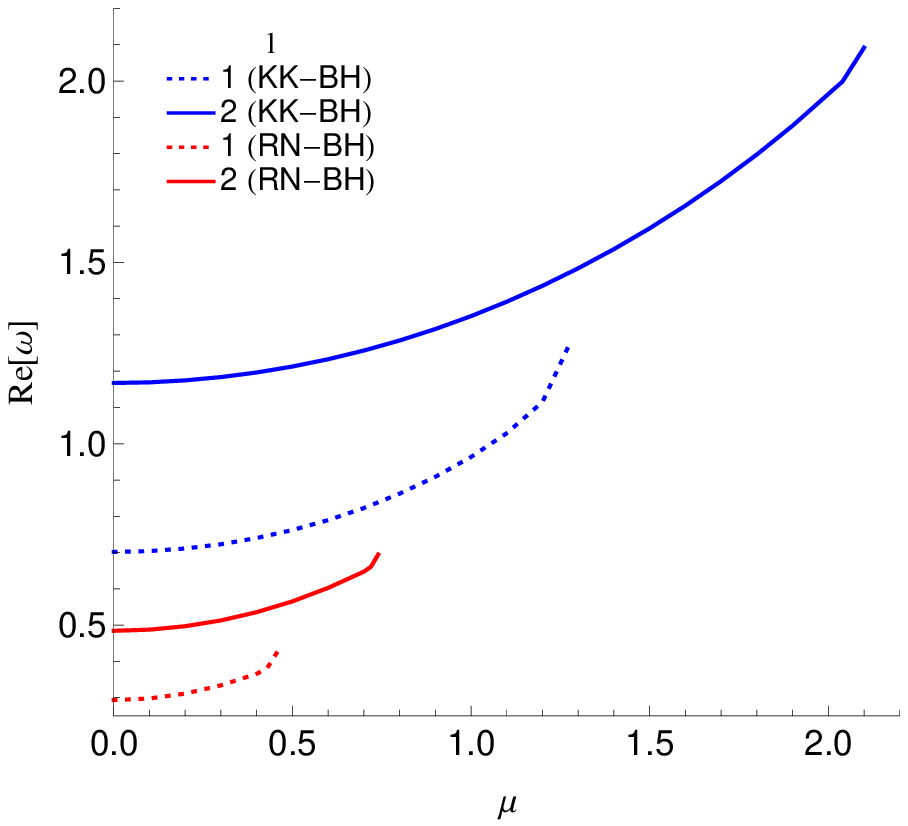} & \epsfxsize=7.5cm %
\epsffile{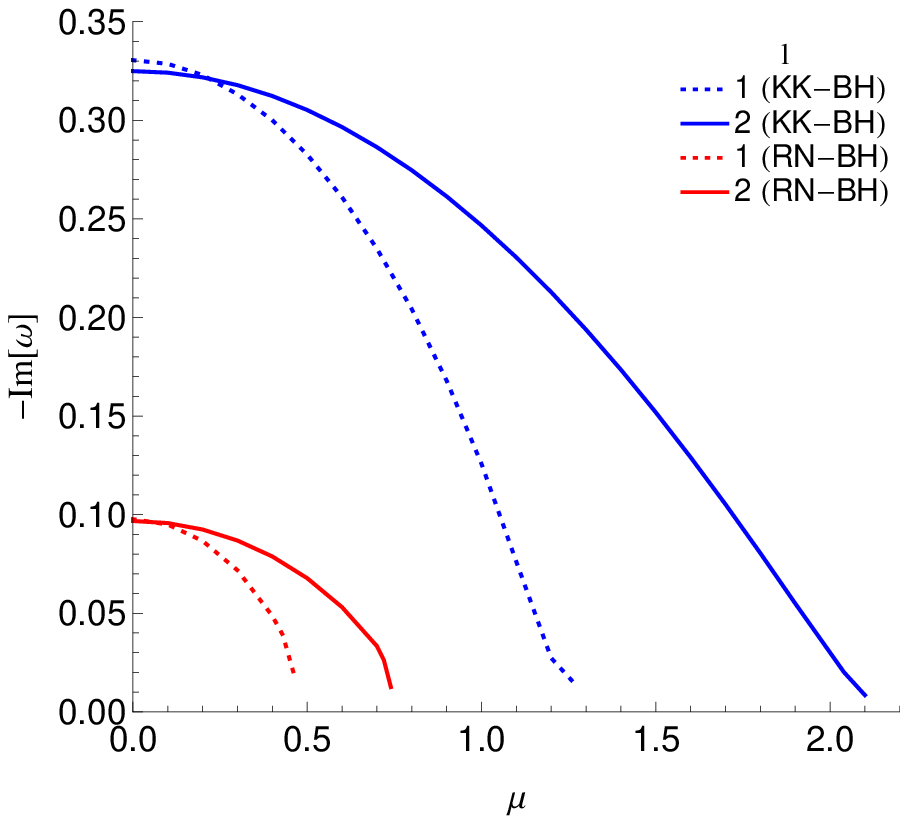} &
\end{array}
$%
\caption{The QN modes for $m=1$ and $Q=0.1$. The red curves show the QNMs of
Reissner-Nordstr\"{o}m BH for $q_{RN}=0.1$. The imaginary part tends to zero
as the field mass increases, hence QNMs disappear and the quasi-resonances
dominate.}
\label{muFig}
\end{figure}

Figure \ref{muFig} shows the behavior of QN frequencies with increasing in $%
\mu $ ranges from zero to $1.27$\ ($2.1$) for $l=1$\ ($l=2$). The red curves
show the QNMs of the Reissner-Nordstr\"{o}m BH and have been plotted for
comparison. As $\mu $ increases, the real part of frequencies increases too
whereas the imaginary part tends to zero, hence QNMs disappear and the
quasi-resonances dominate.

\section{Closing Remark}\label{sec5}

In conclusion, we investigated the critical behavior of rotating KK BH
solution in the presence of Maxwell electrodynamics. Firstly, the conserved
and thermodynamic quantities are introduced. The first law is established
and limited behavior of thermodynamic quantities are considered. The
interesting point was the non-vanishing values of entropy in the presence of
large radius while the temperature is zero in this condition. Electric and
magnetic potential are constant when $x_+\to\infty$.

In order to study the critical behavior of the solution, we considered the
treatment of heat capacity. It is a famous approach as the stability in
canonical ensemble which expresses that stable BHs have positive heat
capacity. In order to specify thermodynamically stable condition of the
solution, we used the numerical method.

Based on the plots we observed that the BH solution should have small event
horizon radius since large event horizon radii caused negative heat
capacity, even though the temperature is positive. To understand the impact
of the charges on the heat capacity we plotted three figures and increased
the value of magnetic parameter when electric charge parameter is fixed. Two
divergencies are observed that the sign of heat capacity is changed by them,
and therefore, we could not call them the first order phase transition. It
may be interpreted as the Davis phase transition.

In addition, we have considered a massive scalar perturbation
minimally coupled to the background geometry of $4$-dimensional
static KK BHs. First, we converted the KK background to a
spherically symmetric line element to obtain a second-order radial
master wave equation, and then calculated the QN modes by
employing the WKB approximation of various orders. From the error
estimation of the WKB formula, we found that the minimum error of
this approximation depends on the charge value $Q$ such that the
best order of the WKB formula for $Q=0.1$ was $7$th-order whereas
the QN frequency had the best accuracy with the help of
$5$th-order for $Q=0.2$. The oscillations increased and the modes
lived longer as the charge $Q$\ decreased.

Besides, the anomalous decay rate of the quasinormal modes spectrum has been
investigated by using the sixth-order WKB formula and we observed that the
curves crossed over at a special critical mass $\tilde{\mu}$. Thus, the KK
BHs in asymptotically flat spacetime had the anomalous decay rate in its QNM
spectrum. We also found that the charge parameter $Q$\ affects the critical
mass and $\tilde{\mu}$\ increases with decreasing in $Q$.

Moreover, the quasi-resonance modes of our BH case study have been
investigated by employing the averaging of Pad\'{e} approximations. It was
shown that, unlike the Schwarzschild case in which the SD formula of
averaging Pad\'{e} approximates of $13$th order is minimal for all values of
the field mass, the minimal SD changed based on the scalar mass for the KK
BHs so that the higher orders have led to minimal SD for low-$\mu $ values
and the lower orders led to minimal SD for high-$\mu$ values.

\appendix

\section{Possible relation between thermal stability and QNMs}

In this appendix, we are going to investigate a possible relation
between thermal stability and QNMs nearby the divergence point of
the heat capacity. Indeed, it is quite interesting to find a
relation between thermal stability and dynamical stability of
black holes and such a connection was suggested in \cite{JingPan}
for the Reissner--Nordstr\"{o}m black holes. In the
aforementioned paper, it is shown that the QNMs of the Reissner--Nordstr\"{o}%
m solutions start to get a spiral-like shape in the complex
$\omega $ plane and both the real and imaginary parts become the
oscillatory functions of the charge whenever the real part of the
QN frequencies arrives at its maximum at the divergence point of
the heat capacity. However, Berti and Cardoso have shown that this
relation is probably due to a numerical coincidence and the
conjectured correspondence does not straightforwardly generalize
to other black hole solutions \cite{BertiCardoso}. In addition, a
similar relationship has been also found between the van der
Waals-like small-large black hole phase transition and QNMs, but
again for the Reissner--Nordstr\"{o}m black holes \cite{Wang}.
Besides, a spiral-like shape in the complex $\omega $ plane has
been reported for the Schwarzschild black holes in the presence of
Quintessence field for the fundamental QNMs and high  multipole
number when the QNMs meet the divergence of the heat capacity
\cite{Tharanath}. However, we should note that these solutions are
very similar to the Reissner--Nordstr\"{o}m black holes as well,
and therefore, observing such a relation was expected.

\begin{figure}[tbp]
$%
\begin{array}{ccc}
\epsfxsize=7.5cm \epsffile{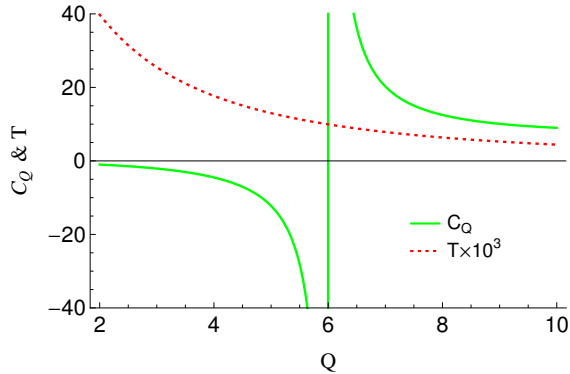}  &
\end{array}
$%
\caption{The heat capacity and temperature versus $Q$ for the
solutions (\ref{SECMetric}). The heat capacity has a divergence at
$Q=6m$ and this figure is plotted for $m=1$.} \label{cStatic}
\end{figure}
\begin{figure}[tbp]
$%
\begin{array}{ccc}
\epsfxsize=7.5cm \epsffile{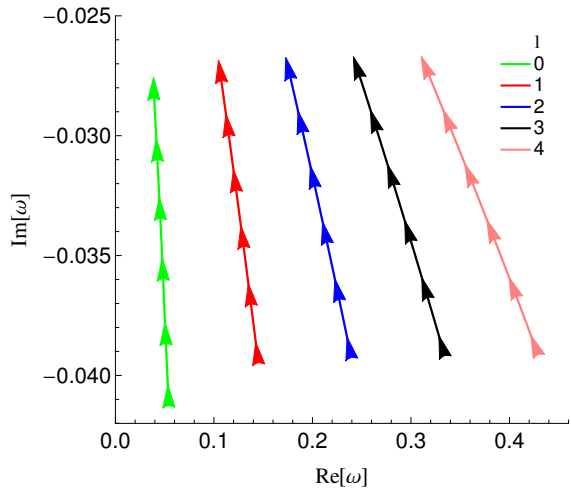} &
\end{array}
$%
\caption{The fundamental QN modes in the complex $\omega $ plane
for different values of the multipole number. The charge $Q$
starts from $Q=5$ and ends at $Q=7$ in each curve along with the
arrows.} \label{IR}
\end{figure}

Now, we search for a spiral-like shape in the complex $\omega $
plane for the static black hole solutions given in
(\ref{SECMetric}) at the divergence point of the heat capacity. It
is straightforward to show that the heat capacity of this solution
has a divergence at $Q=6m$, as it is shown in Fig. \ref{cStatic}
for the special case $m=1$. Therefore, the QNMs for the
fundamental mode around $Q=6$\ have been calculated and the
results are illustrated through Fig. \ref{IR} in the complex
$\omega $ plane. As one can see from this figure, there is no
spiral-like shape around the divergence point of the heat capacity
located at $Q=6$, and thus, we cannot observe a connection between
QNMs and thermal stability for these black hole solutions.
However, although this connection is not observed in Fig.
\ref{IR}, it maybe appear for other choices of overtone and
multipole numbers which is needed to be investigated in more
details via more powerful numerical methods than the WKB one. We
shall address this point in future work.


\acknowledgments{SHH, SH and MM thank Shiraz University Research
Council.}



\begin{thebibliography}{99}
\bibitem{KK} A. Chodos and S.L. Detweiler,
Gen. Rel. Grav. \textbf{14}, 879 (1982);\newline
P. Dobiasch and D. Maison,
Gen. Rel. Grav. \textbf{14}, 231 (1982);\newline
D. Rasheed,  
Nucl. Phys. B \textbf{454}, 379 (1995).

\bibitem{KK1} G.W. Gibbons and D.L. Wiltshire.
Annals of Physics \textbf{167(1)}, 201 (1986).

\bibitem{KK2} G.T. Horowitz and T. Wiseman,
arXiv:1107.5563 (2011);\newline
G.T. Horowitz eds., BHs in Higher Dimensions, Cambridge University Press,
cambridge U.K. (2012).

\bibitem{KK5} P. Dobiasch and D. Maison,
Gen. Rel. Grav. \textbf{14}, 231 (1982).

\bibitem{KK7} G. Landi, N. Viet and K. Wali,
Phys. Lett. B \textbf{326}, 45 (1994).

\bibitem{KK4} F. Larsen,  
Nucl. Phys. B, \textbf{575}, 211 (2000).

\bibitem{KK6} J. Park,
Class. Quant. Grav. \textbf{15}, 775 (1998).

\bibitem{phase1} P. C. W. Davies,  
Proc. R. Soc. Lond. A \textbf{353}, 499 (1977);\newline
P. C. W. Davies,  
Rep. Prog. Phys. \textbf{41}, 1313 (1978);\newline
P. C. W. Davies,
Class. Quant. Grav. \textbf{6}, 1909 (1989).

\bibitem{phase2} P. Hut,  
Mon. Not. R. Astr. Soc. \textbf{180}, 379 (1977).


%
%


\bibitem{Kokkotas} K. D. Kokkotas and B. G. Schmidt, Living Rev. Rel.
\textbf{2}, 2 (1999).

\bibitem{BertiRev} E. Berti, V. Cardoso and A. O. Starinets, Class. Quant.
Grav. \textbf{26}, 163001 (2009).

\bibitem{GWsupMass} E. Berti, V. Cardoso and C. M. Will, Phys. Rev. D
\textbf{73}, 064030 (2006).

\bibitem{Abbott1} B. P. Abbott et al. [LIGO Scientific and Virgo
Collaborations], Phys. Rev. Lett. \textbf{116}, 061102 (2016).

\bibitem{Cheung} C. Cheung, A. L. Fitzpatrick, J. Kaplan, L. Senatore and P.
Creminelli, JHEP \textbf{03}, 014 (2008).

\bibitem{Gubitosi} G. Gubitosi, F. Piazza and F. Vernizzi, JCAP \textbf{02},
032 (2013).

\bibitem{Hu} W. Hu, R. Barkana and A. Gruzinov, Phys. Rev. Lett. \textbf{85}%
, 1158 (2000).

\bibitem{Arvanitaki} A. Arvanitaki, S. Dimopoulos, S. Dubovsky, N. Kaloper,
and J. M. Russell, Phys. Rev. D \textbf{81}, 123530 (2010).

\bibitem{Metsaev} R. Metsaev and A. Tseytlin, Nucl. Phys. B \textbf{293},
385 (1987).

\bibitem{Herdeiro} C. A. R. Herdeiro and E. Radu, Int. J. Mod. Phys. D
\textbf{24}, 1542014 (2015).

\bibitem{Silva} H. O. Silva, J. Sakstein, L. Gualtieri, T. P. Sotiriou, and
E. Berti, Phys. Rev. Lett. \textbf{120}, 131104 (2018).

\bibitem{Brito} R. Brito, V. Cardoso and P. Pani, Lect. Notes Phys. \textbf{%
906}, 1 (2015).

\bibitem{Clough} K. Clough, P. G. Ferreira and M. Lagos, Phys. Rev. D
\textbf{100}, 063014 (2019).

\bibitem{Tattersall} O. J. Tattersall and P. G. Ferreira, Phys. Rev. D
\textbf{97}, 104047 (2018).

\bibitem{Dalang} C. Dalang, P. Fleury and L. Lombriser, Phys. Rev. D \textbf{%
103}, 064075 (2021).

\bibitem{Maselli} A. Maselli, N. Franchini, L. Gualtieri, T. P. Sotiriou, S.
Barsanti and P. Pani,  
arXiv:2106.11325 (2021).

\bibitem{SchwQRMs} R. A. Konoplya and A. Zhidenko, Phys. Lett. B \textbf{609}%
, 377 (2005).

\bibitem{higherSchwQRM} A. Zhidenko, Phys. Rev. D \textbf{74}, 064017 (2006).

\bibitem{ADR} M. Lagos, P. G. Ferreira and O. J. Tattersall, Phys. Rev. D
\textbf{101}, 084018 (2020).

\bibitem{RN-QRM} A. Ohashi and M. A. Sakagami, Class. Quant. Grav. \textbf{21%
}, 3973 (2004).

\bibitem{Hod-rnQRM} S. Hod, Phys. Lett. B \textbf{761}, 53 (2016).

\bibitem{ADRrn} R. D. B. Fontana, P. A. Gonzalez, E. Papantonopoulos and Y.
Vasquez, Phys. Rev. D \textbf{103}, 064005 (2021).

\bibitem{magnetizedSchw-QRM} C. Wu and R. Xu, Eur. Phys. J. C \textbf{75},
391 (2015).

\bibitem{KerrQRM} R. A. Konoplya and A. Zhidenko, Phys. Rev. D \textbf{73},
124040 (2006).

\bibitem{EinWeylGravityQRM} A. F. Zinhailo, Eur. Phys. J. C, \textbf{78},
992 (2018).

\bibitem{MehrabWeylNE} M. Momennia and S. H. Hendi, Phys. Rev. D \textbf{99}%
, 124025 (2019).

\bibitem{MehrabWeylGPs} M. Momennia and S. H. Hendi, Eur. Phys. J. C \textbf{%
80}, 505 (2020).

\bibitem{Grigoris} A. Rincon and G. Panotopoulos, Phys. Rev. D \textbf{97},
024027 (2018).

\bibitem{KK3} D. Rasheed,
Nucl. Phys. B \textbf{454}, 379 (1995).


\bibitem{KK9} M. Azreg-Ainou, M. Jamil and K. Lin,
Chinese Physics C \textbf{44(6)}, 065101 (2020).

\bibitem{Xray} J. Zhu, et al.,
Eur. Phys. J. C \textbf{80}, 622 (2020).

\bibitem{Ghasemi} M. Ghasemi-Nodehi, M. Azreg-Ainou, K. Jusufi and M. Jamil,
Phys. Rev. D \textbf{102}, 104032 (2020).

\bibitem{KK10} R.G. Cai, L. M. Cao and N. Ohta,
Phys. Lett. B \textbf{639} 354 (2006).


\bibitem{hc1} P. C. W. Davies, 
Proc. R. Soc. Lond. A \textbf{353}, 499 (1977);\newline
P. C. W. Davies, 
Rep. Prog. Phys. \textbf{41}, 1313 (1978).

\bibitem{Schutz} B. F. Schutz and C. M. Will, Astrophys. J. Lett. \textbf{291%
}, L33 (1985).

\bibitem{IyerWill} S. Iyer and C. M. Will, Phys. Rev. D \textbf{35}, 3621
(1987).

\bibitem{Konoplya6th} R. A. Konoplya, Phys. Rev. D \textbf{68}, 024018
(2003).

\bibitem{Matyjasek13th} J. Matyjasek and M. Opala, Phys. Rev. D \textbf{96},
024011 (2017).

\bibitem{wkbError} R. A. Konoplya, A. Zhidenko and A. F. Zinhailo, Class.
Quant. Grav. \textbf{36}, 155002 (2019).

\bibitem{WeylGravityQRM} M. Momennia, S. H. Hendi and F. Soltani Bidgoli,
Phys. Lett. B \textbf{813}, 136028 (2021).

\bibitem{WormQRM} M. S. Churilova, R. A. Konoplya and A. Zhidenko, Phys.
Lett. B \textbf{802}, 135207 (2020).

\bibitem{SchwDS-QRMs} R. A. Konoplya, Phys. Rev. D \textbf{73}, 024009
(2006).

\bibitem{JingPan} J. Jing and Q. Pan, Phys. Lett. B \textbf{660}, 13 (2008).

\bibitem{BertiCardoso} E. Berti and V. Cardoso, Phys. Rev. D \textbf{77},
087501 (2008).

\bibitem{Wang} Y. Liu, D. C. Zou and B. Wang, JHEP \textbf{09}, 179 (2014).

\bibitem{Tharanath} R. Tharanath, N. Varghese and V. C. Kuriakose, Mod.
Phys. Lett. A \textbf{29}, 1450057 (2014).
\end{thebibliography}
\end{document}